\documentclass{article}

\usepackage{arxiv}

\usepackage[utf8]{inputenc} 
\usepackage[T1]{fontenc}    
\usepackage{hyperref}       
\usepackage{url}            
\usepackage{booktabs}       
\usepackage{amsfonts}       
\usepackage{nicefrac}       
\usepackage{microtype}      
\usepackage{lipsum}		
\usepackage{graphicx}
\usepackage{natbib}
\usepackage{doi}
\usepackage{amsmath}

\title{Modelling Expert Cognition Beyond Behaviour: Towards Interpretation, Tension, and Value Structures}

\author{ 
	\href{https://orcid.org/0009-0004-1760-0149}{\includegraphics[scale=0.06]{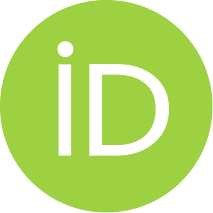}\hspace{1mm}Annie Yuan} \\
	School of Computer Science\\
	The University of Sydney\\
	NSW, 2006, Australia \\
	\texttt{annie.yuan@sydney.edu.au} \\
}

\renewcommand{\shorttitle}{Modelling Expert Cognition Beyond Behaviour}

\hypersetup{
pdftitle={A template for the arxiv style},
pdfsubject={q-bio.NC, q-bio.QM},
pdfauthor={David S.~Hippocampus, Elias D.~Striatum},
pdfkeywords={First keyword, Second keyword, More},
}

\begin{document}
\maketitle

\begin{abstract}
Existing computational models of expertise primarily focus on observable behaviour or decision outcomes, failing to capture the internal cognitive structures that generate expert reasoning. In this work, we introduce the Expert Identity Cognition Model (EICM), a three-layer framework for modelling expert cognition beyond behaviour.
EICM conceptualises expert cognition as an identity-structured process operating within situational constraints, where constraints are interpreted through internal tensions arising from competing identity commitments and stabilised into value structures that guide action.
Unlike behaviour-centric or constraint-driven approaches, EICM positions tension as the central cognitive mechanism connecting world structure and decision formation. We argue that expert cognition is not merely behavioural adaptation under constraints but an identity-structured negotiation process that produces stable judgement patterns across contexts.
The framework provides a new perspective for modelling tacit knowledge, expert judgement, and cognitive consistency in domains including professional practice, cultural expertise, and design reasoning.

\end{abstract}


\section{Introduction}

modelling expert cognition remains a central challenge in artificial intelligence, cognitive science, and human-centred computing. Many computational approaches to expertise model expert behaviour through demonstrations, trajectories, reward structures, preferences, or optimised decision outputs \cite{hussein2017imitation,sutton1998reinforcement,christiano2017deep}. These approaches have achieved substantial success in domains where expertise can be approximated through repeatable behavioural patterns or performance-based optimisation. However, many forms of expertise---particularly those involving tacit knowledge, cultural judgement, creativity, and long-term professional reasoning---cannot be fully explained through behaviour alone.

Research on expertise and professional judgement has long emphasised that expert reasoning is often tacit, situated, and difficult to reduce to explicit rules or procedures \cite{polanyi1966logic,dreyfus1986mind,schon2017reflective}. Similarly, work on situated and distributed cognition shows that cognition unfolds through interaction with material, social, and environmental conditions rather than through isolated internal computation alone \cite{suchman1987plans,hollan2000distributed}. These traditions suggest that expert action is not merely an output to be imitated, but the visible expression of deeper interpretive processes.

A key challenge, therefore, is to explain how experts interpret constraints, negotiate internal conflicts, and stabilise judgement across changing contexts. Experts operating under similar external conditions may produce different judgements, priorities, and decision strategies. Conversely, similar observable actions may emerge from different internal reasoning structures. This suggests a need for models that move beyond behavioural reproduction towardss reconstruction of the latent cognitive organisation underlying expert judgement.

We argue that expert cognition is not fundamentally behaviour-driven, but identity-structured. Experts do not merely respond to external constraints. Instead, they interpret constraints through internal tensions emerging from competing identity commitments. These tensions are then stabilised into value structures that guide action. This argument builds on work showing that identity and self-concept shape motivation, interpretation, and action \cite{markus1987dynamic,oyserman2009identity}, and that values provide relatively stable structures for evaluation and prioritisation \cite{schwartz1992universals}.

To address this gap, we introduce the \emph{Expert Identity Cognition Model} (EICM), a three-layer cognitive architecture composed of \textbf{Constraint} as situational structure, \textbf{Tension} as identity conflict structure, and \textbf{Value} as decision structure. EICM models expertise as a transformation process in which external conditions are interpreted through identity tensions and stabilised into value systems.

In summary, we propose EICM as an identity-structured framework for expert cognition that moves beyond behaviour modelling to formalise how constraints are interpreted, tensions are formed, and values are stabilised.

This paper makes four primary contributions:

\begin{itemize}
    \item We introduce EICM, a three-layer framework that models expert cognition through Constraint, Tension, and Value.
    \item We conceptualise identity tension as the mediating mechanism through which experts interpret situational constraints.
    \item We define values as stabilised decision principles rather than reward functions, abstract preferences, or static moral claims.
    \item We demonstrate how this framework redirects expert modelling from behavioural imitation towards identity-structured cognitive reconstruction.
\end{itemize}

\section{Related Work}

\subsection{Behaviour-Centric and Preference-Based Expert modelling}

A dominant paradigm in AI models expertise through observable behaviour. In imitation learning, expert demonstrations are used to learn mappings between states and actions, with the goal of reproducing expert-like performance in similar task environments \cite{hussein2017imitation}. Reinforcement learning similarly models decision-making through interaction with an environment, where agents learn policies that maximize expected reward over time \cite{sutton1998reinforcement}. More recent preference-based approaches extend this paradigm by using human feedback to infer reward models or preference signals, allowing systems to optimise behaviour according to human judgements rather than manually specified objectives \cite{christiano2017deep}. This logic has also become central to contemporary language model alignment, where models are trained to follow human instructions through feedback-based optimisation procedures \cite{ouyang2022training}.

These approaches have been highly influential because they provide scalable mechanisms for learning competent behaviour from demonstrations, rewards, or feedback. However, they tend to operationalise expertise through externalised outputs: actions, trajectories, preferences, or reward-aligned responses. As a result, expert cognition is often treated as behaviourally inferable, with internal reasoning structures abstracted away or represented only indirectly through learned policies or preference models.

EICM departs from this behaviour-centric framing. Rather than treating observable behaviour as the primary locus of expertise, EICM conceptualises behaviour as the downstream manifestation of internal cognitive organisation. From this perspective, demonstrations and preferences may reveal what an expert does or chooses, but they do not necessarily explain how constraints are interpreted, how internal tensions are negotiated, or how stable value structures are formed. This distinction is particularly important in domains involving tacit judgement, cultural reasoning, professional identity, or creative expertise, where similar observable actions may emerge from different cognitive structures, and similar constraints may produce divergent expert judgements.

Accordingly, EICM positions expert modelling not as the reproduction of behaviour alone, but as the reconstruction of the identity-mediated interpretive structures that generate behaviour. This shift moves beyond learning state-action mappings or preference approximations towardss modelling the latent cognitive pathway through which constraints become meaningful, tensions are formed, and values are stabilised into action.

\subsection{Constraint, Situated Action, and Distributed Cognition}

A second relevant tradition concerns the role of constraints, situations, and environments in shaping cognition. Simon's account of bounded rationality challenged the view of decision-making as fully optimal reasoning, arguing instead that human cognition operates under limits of information, time, and computational capacity \cite{simon1972theories}. From this perspective, cognition is shaped by the structure of the environment and by the bounded resources available to the decision-maker.

Situated action further complicates the idea that behaviour follows directly from pre-specified plans or internal rules. Suchman argues that action is situated in unfolding interaction with the material and social world, rather than simply executed from abstract plans \cite{suchman1987plans}. Similarly, distributed cognition expands the unit of analysis beyond the individual mind, emphasising how cognition is distributed across people, artefacts, tools, representations, and environments \cite{hollan2000distributed}.

EICM builds on the view that cognition is shaped by constraints and situated environments but interprets constraints as problem spaces that acquire meaning through identity tensions rather than as direct determinants of behaviour. This framing helps explain why experts facing similar external conditions may interpret those conditions differently and arrive at different judgements.

\subsection{Tacit Knowledge and Expert judgement}

A third relevant tradition concerns tacit knowledge and expert judgement. Polanyi's account of tacit inference emphasises that human knowing often exceeds what can be explicitly stated, suggesting that expertise cannot be fully reduced to formal rules or articulated procedures \cite{polanyi1966logic}. This view is echoed in accounts of expert skill, where advanced expertise is understood as increasingly intuitive, situated, and difficult to capture through rule-based representation alone \cite{dreyfus1986mind}. 

Research on professional judgement similarly emphasises that expertise develops through reflective engagement with practice. Sch{\"o}n's notion of reflection-in-action describes how professionals reason within unfolding situations rather than merely applying pre-existing rules \cite{schon2017reflective}. Work on expert intuition and naturalistic decision-making further shows that expert decisions are often made under uncertainty, time pressure, and context-specific constraints, drawing on patterns of experience that may not be consciously articulated \cite{klein2011expert}.

Related work on knowledge creation, practice theory, and situated learning also frames expertise as embedded in social, material, and practice-based contexts. Nonaka and Takeuchi describe the relationship between tacit and explicit knowledge in organisational knowledge creation \cite{nonaka2007knowledge}, while practice-theoretic and situated learning perspectives emphasise that knowledge is developed through participation in ongoing practices and communities \cite{schatzki2001practice,lave1991situated}.

EICM builds on these accounts by treating tacit expertise as an identity-structured process of interpretation and stabilisation. Rather than modelling expert judgement as explicit procedural knowledge alone, EICM focuses on how constraints are interpreted through identity tensions and stabilised into value structures that guide action.

\subsection{Identity, Self, and Value in Cognition}

A fourth relevant tradition concerns identity, self-concept, and values as structures that shape action and meaning. Early work on identities and interaction examined how individuals act through socially organised roles and self-understandings \cite{mccall1966identities}. Identity theory further developed this view by linking identity to role-based meanings, social structure, and patterns of action \cite{stryker2001traditional}. From this perspective, identity is not merely a descriptive label, but a structure that helps organise how individuals understand themselves and act within social contexts.

Psychological accounts of the self also emphasise that self-concept is dynamic rather than fixed. Markus and Wurf describe the self-concept as an active and changing structure that influences information processing, motivation, affect, and behaviour \cite{markus1987dynamic}. Similarly, identity-based motivation theory argues that identities shape action-readiness by making certain goals, interpretations, and strategies feel appropriate or meaningful in context \cite{oyserman2009identity}. These perspectives are important for EICM because they suggest that identity can influence cognition, not only social presentation or personal description.

Values provide another relevant layer for understanding how judgement is organised. Schwartz's theory of basic human values conceptualises values as relatively stable motivational structures that guide evaluation, prioritisation, and action across situations \cite{schwartz1992universals}. This supports the view that expert judgement may involve stable value orientations rather than isolated preferences or one-off decisions.

EICM builds on these traditions by treating identity and value as central to expert cognition. Its contribution is to focus on the tensions that emerge when multiple identity commitments are activated under situational constraints. Rather than modelling identity as a static trait or social category, EICM conceptualises identity as a dynamic tension structure through which constraints are interpreted. Within this framing, values are not simply pre-existing preferences, but stabilised decision principles that emerge through the negotiation of competing identity commitments.

\subsection{Positioning EICM}

Taken together, the reviewed literature shows that expertise has been approached through several complementary lenses. Behaviour-centric and preference-based AI approaches model expertise through actions, demonstrations, rewards, or feedback signals. Work on bounded, situated, and distributed cognition shows that reasoning is shaped by constraints, environments, tools, and social contexts. Research on tacit knowledge and expert judgement shows that expertise often exceeds explicit rules and formal procedures. Finally, work on identity, self, and values suggests that action and judgement are shaped by self-understanding, commitments, and value orientations.

EICM builds on these traditions while proposing a distinct modelling target: the identity-mediated cognitive structures through which experts interpret constraints and stabilise values into action. This positioning is also relevant to human-computer interaction and human-AI systems. Value-sensitive design has long argued that human values should be accounted for throughout the design of information systems \cite{friedman2013value}, with later work providing methodological guidance for investigating and integrating values in HCI research and design practice \cite{friedman2017survey}. EICM is complementary to this tradition but shifts the focus from designing for values to modelling how expert values are cognitively formed, negotiated, and stabilised through identity tensions.

EICM also connects to human-AI interaction research concerned with interpretability, user understanding, and mental models. Recent work on mental models in human-AI interaction highlights the importance of understanding how people conceptualise, reason about, and interact with AI systems \cite{sanchez2026mental}. EICM extends this concern in the direction of expert modelling: rather than asking only how users understand AI systems, it asks how AI systems might represent the interpretive structures through which experts understand constraints and make judgements.

Finally, EICM is relevant to emerging HCI and AI work on tacit knowledge in computational practice. For example, studies of machine learning data work have shown that important forms of expertise often remain tacit, situated, and difficult to formalise directly \cite{cha2023unlocking}. EICM contributes a framework for representing such expertise not merely as undocumented practice or implicit skill, but as an identity-structured process linking constraints, tensions, values, and action.

Accordingly, EICM reframes expert cognition modelling as more than behavioural reproduction or preference approximation. Its central contribution is to position \emph{identity tension} as a core cognitive mechanism through which constraints are interpreted, values are stabilised, and expert action becomes intelligible. This offers a foundation for human-centred AI systems that aim not only to imitate expert behaviour, but to reconstruct and explain the latent cognitive organisation underlying expert judgement.

\section{Expert Identity Cognition Model}

\subsection{Definition}

We define expert cognition as an identity-structured process operating within situational constraints, in which internal tensions between identity commitments are resolved into stable value structures that guide action.

Based on this formulation, we introduce the \emph{Expert Identity Cognition Model} (EICM), a three-layer cognitive architecture composed of \emph{Constraint}, \emph{Tension}, and \emph{Value}. EICM models expertise not as direct behavioural adaptation, but as a layered process of interpretation, internal negotiation, and value stabilisation.

Unlike conventional behaviour-centric approaches, EICM assumes that cognition cannot be inferred solely from observable action. Instead, expert reasoning emerges through the interaction between external situational structures and internal identity commitments.

Within the framework, constraints define the external condition space of cognition; tensions organise interpretive conflict between identity commitments; and values stabilise decision principles that ultimately generate observable action. Accordingly, EICM positions behaviour not as cognition itself, but as the downstream expression of stabilised internal cognitive organisation.

\subsection{Constraint}

Constraint refers to the structured situational conditions within which expert cognition operates. These constraints include, but are not limited to:

\begin{itemize}
    \item physical limitations;
    \item institutional regulations;
    \item task-specific requirements;
    \item informational incompleteness;
    \item temporal uncertainty;
    \item social expectations; and
    \item resource scarcity.
\end{itemize}

In EICM, constraints are not treated as deterministic drivers of behaviour. Instead, they define the external problem space that cognition must interpret and negotiate.

This distinction is critical. Traditional constraint-driven models often assume that behaviour emerges directly from environmental pressures or optimisation requirements. However, experts facing highly similar constraints frequently arrive at different judgements, priorities, and strategies.

EICM explains this divergence by arguing that constraints do not possess intrinsic cognitive meaning independent of interpretation. Rather, constraints become cognitively significant only when interpreted through identity structures. Thus, the function of the Constraint layer is not to generate action directly, but to establish the situational structure within which cognitive interpretation becomes necessary.

Formally, the constraint space can be represented as:

\begin{equation}
C = \{c_1, c_2, \ldots, c_n\}
\end{equation}

where each $c_i$ represents a structured situational condition influencing cognition.

Importantly, the model assumes that constraints are partially interpretable rather than fully objective. The same external condition may acquire different meanings depending on the identity commitments through which it is perceived.

\subsection{Tension}

Tension constitutes the central cognitive mechanism of EICM. We define tension as the structured internal conflict emerging between competing identity commitments under situational constraints.

Unlike conventional models that interpret conflict primarily as uncertainty, ambiguity, or optimisation difficulty, EICM conceptualises tension as identity-structured cognitive friction. Tension arises not merely because decisions are difficult, but because experts simultaneously maintain multiple commitments that cannot be fully satisfied under existing constraints.

Examples include conflicts between:

\begin{itemize}
    \item tradition and innovation;
    \item precision and expression;
    \item efficiency and craftsmanship;
    \item responsibility and creativity; and
    \item stability and transformation.
\end{itemize}

Within EICM, tension functions as the interpretive engine of cognition. External constraints do not directly produce reasoning; instead, reasoning emerges through the negotiation of competing identity commitments activated by those constraints.

This formulation explains why identical situations may generate fundamentally different judgements across experts. Two experts may observe the same external condition while interpreting its significance through entirely different identity structures. Accordingly, EICM positions tension as the minimal cognitive unit of expert reasoning.

The tension structure can be formally represented as:

\begin{equation}
T = f(C, I)
\end{equation}

where $C$ represents situational constraints, $I$ represents identity commitments, and $T$ represents the resulting tension structure.

Identity commitments themselves may be represented as:

\begin{equation}
I = \{i_1, i_2, \ldots, i_k\}
\end{equation}

where each $i_k$ corresponds to a relatively stable professional, cultural, ethical, or epistemic commitment.

Tension therefore functions as the mediation layer between world structure and cognitive interpretation.

\subsection{Value}

Value refers to stabilised decision structures emerging through the negotiation and partial resolution of identity tensions.

In EICM, values are not interpreted as abstract moral principles, static preferences, or reward functions. Instead, they represent relatively stable cognitive priorities that organise judgement consistency across contexts.

Examples include:

\begin{itemize}
    \item prioritisation of craftsmanship integrity;
    \item preservation of symbolic meaning;
    \item long-term quality orientation;
    \item expressive authenticity; and
    \item professional responsibility.
\end{itemize}

Importantly, values are not assumed to exist independently of tension. Rather, they emerge through repeated stabilisation processes in which experts negotiate competing commitments under constraints. This stabilisation process explains why experienced experts frequently demonstrate consistent decision patterns even across changing environments.

Within EICM, value structures function as cognitive organisers that transform internal negotiation into actionable judgement. Formally:

\begin{equation}
V = g(T)
\end{equation}

where $T$ represents identity tension structures and $V$ represents stabilised value structures.

Observable action is subsequently generated through these value structures rather than directly from constraints themselves. Thus, value operates as the intermediate stabilising layer between cognition and behaviour.

\subsection{Formalisation}

The complete EICM process can be expressed as a layered transformation structure:

\begin{align}
T &= f(C, I) \\
V &= g(T) \\
A &= h(V)
\end{align}

where $C$ denotes situational constraints, $I$ denotes identity commitments, $T$ denotes tension structures, $V$ denotes stabilised value structures, and $A$ denotes observable action.

Combining these formulations yields:

\begin{equation}
A = h(g(f(C, I)))
\end{equation}

This representation formalises expert cognition as a mediated interpretive process rather than direct behavioural response. The formulation further implies that behaviour alone is insufficient for reconstructing expert cognition, since action represents only the downstream manifestation of stabilised value structures.

Accordingly, EICM reframes expertise as identity-mediated cognitive organisation rather than behavioural competence alone.

\section{Identity-Structured Cognition}

\begin{figure}[h]
    \centering
    \includegraphics[width=0.95\linewidth]{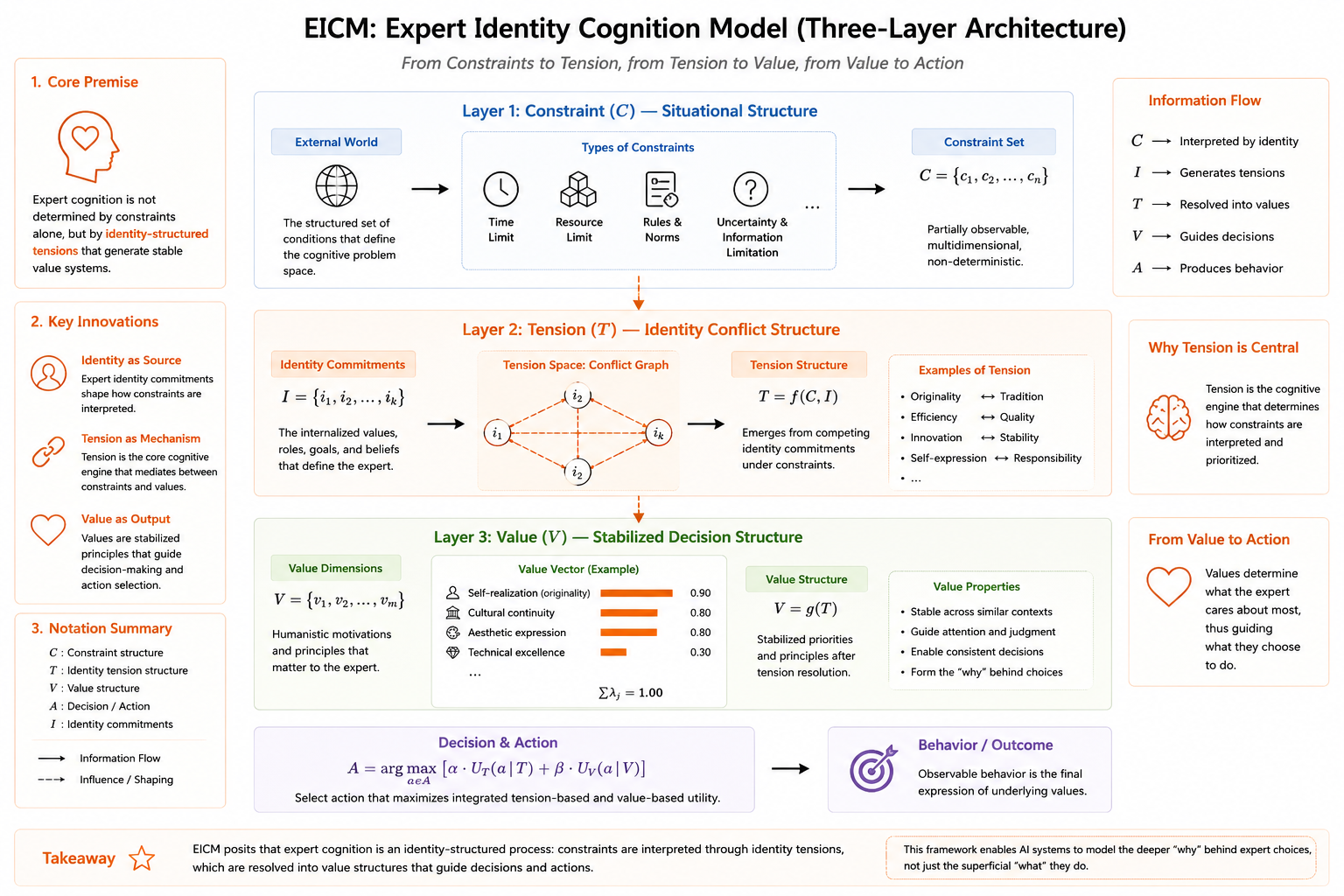}
    \caption{Architecture of the Expert Identity Cognition Model (EICM). The model represents expert cognition as a layered process in which situational constraints are interpreted through identity tensions, stabilised into value structures, and expressed through observable action.}
    \label{fig:eicm-architecture}
\end{figure}

This section develops the theoretical structure of EICM as an identity-structured account of expert cognition. Figure~\ref{fig:eicm-architecture} illustrates the model as a layered cognitive process in which situational constraints are interpreted through identity tensions, stabilised into value structures, and ultimately expressed through observable action.

\subsection{Identity-Structured Expert Cognition}

Figure~\ref{fig:eicm-architecture} presents the central theoretical architecture of the \emph{Expert Identity Cognition Model} (EICM). The figure illustrates a fundamental shift from behaviour-centric models of expertise towards an identity-structured account of cognition.

Traditional computational approaches frequently model expertise through observable behaviour, optimisation performance, or decision outputs. Within such approaches, cognition is implicitly treated as a direct response to external constraints. EICM challenges this assumption by proposing that constraints alone do not generate expert reasoning. Instead, expert cognition emerges through the mediation of internal identity tensions.

\subsection{Identity-Structured Expert Cognition}

Figure~\ref{fig:eicm-architecture} presents the central theoretical architecture of the \emph{Expert Identity Cognition Model} (EICM). The figure illustrates a fundamental shift from behaviour-centric models of expertise towards an identity-structured account of cognition.

Traditional computational approaches frequently model expertise through observable behaviour, optimisation performance, or decision outputs. Within such approaches, cognition is implicitly treated as a direct response to external constraints. EICM challenges this assumption by proposing that constraints alone do not generate expert reasoning. Instead, expert cognition emerges through the mediation of internal identity tensions.

The first layer of the model, \emph{Constraint}, represents the structured situational conditions within which cognition occurs. These conditions include task requirements, institutional rules, resource limitations, and uncertainty structures. In EICM, constraints define the external problem space but do not determine cognition directly.

The central contribution of the framework lies in the second layer, \emph{Tension}. EICM positions tension as the primary cognitive mechanism through which experts interpret constraints. Unlike conventional notions of conflict as optimisation difficulty or decision ambiguity, tension is modeled here as identity-structured cognitive friction emerging between competing identity commitments. Importantly, identical constraints may produce entirely different reasoning trajectories depending on how experts negotiate these tensions internally.

This formulation explains why expert cognition often cannot be inferred from behaviour alone. Two experts operating under similar external conditions may produce different judgements not because of informational differences, but because they stabilise different identity tensions. EICM therefore reframes expertise as a structured process of internal negotiation rather than behavioural reaction.

The third layer, \emph{Value}, represents the stabilisation of decision principles emerging from tension resolution. Values within EICM are not treated as abstract preferences or reward functions, but as relatively stable cognitive priorities that organise consistent judgement and action. Observable behaviour is therefore interpreted as the downstream manifestation of stabilised value structures rather than the primary locus of cognition itself.

The directional structure of Figure~\ref{fig:eicm-architecture} reflects this layered transformation process. Constraints define the situational condition space; tensions organise cognitive interpretation through identity conflict; values stabilise decision formation; and behaviour emerges as the external expression of this internal cognitive organisation.

Through this formulation, EICM advances a new perspective on expert modelling beyond behavioural imitation. The framework positions identity tension as the central bridge between external conditions and expert decision formation, providing a foundation for modelling tacit knowledge, professional reasoning, and long-term judgement consistency across domains.

\subsection{Tension as the Core Cognitive Mechanism}

EICM positions tension as the central mechanism of expert cognition. This claim represents a significant departure from conventional approaches in which cognition is modeled primarily through optimisation, rule application, or behavioural adaptation.

Within EICM, cognition emerges through the negotiation of incompatible or partially conflicting identity commitments activated under constraints. Importantly, tension is not equivalent to uncertainty or decision difficulty alone. Two situations may involve equal uncertainty while producing entirely different cognitive dynamics depending on the identity structure of the expert involved.

Tension therefore explains how experts interpret the same conditions differently despite operating under comparable informational environments. For example, under identical institutional pressure, one expert may prioritise preservation of traditional integrity, while another prioritises adaptation and innovation. The divergence emerges not from the constraint itself, but from the identity tensions through which the constraint is interpreted.

This formulation enables EICM to account for tacit reasoning, long-term stylistic consistency, and professional judgement patterns that cannot be fully reduced to observable behaviour. Accordingly, EICM proposes that identity tension, rather than action, is the minimal reconstructable unit of expert cognition.

\subsection{Beyond Behaviour modelling}

A central implication of EICM is that expert cognition cannot be adequately modeled through behaviour alone. Most existing AI approaches to expertise rely on observable actions, demonstrations, or preference outputs as proxies for cognition. While effective in highly structured domains, such approaches remain insufficient for modelling tacit judgement, cultural expertise, or identity-dependent reasoning processes.

Behavioural outputs alone cannot explain:

\begin{itemize}
    \item why experts facing similar conditions diverge in judgement;
    \item why consistent reasoning patterns persist across contexts;
    \item how professional identities shape interpretation; or
    \item how tacit values stabilise over time.
\end{itemize}

EICM addresses this limitation by relocating the primary locus of cognition from behaviour to identity-mediated tension structures. Within this perspective, observable action becomes the downstream manifestation of stabilised cognitive organisation rather than the direct object of modelling itself.

This shift has significant implications for AI expert modelling, tacit knowledge reconstruction, and human-centred intelligent systems. Instead of reproducing behaviour alone, future systems may require the reconstruction of internal identity tensions and value stabilisation processes underlying expert reasoning.

Accordingly, EICM proposes a transition from behaviour imitation towards identity-structured cognitive reconstruction as a new direction for modelling expertise.

\section{Illustrative Case Study: Persona-Induced Decision Divergence Under Identical Constraints}

A central claim of EICM is that expert cognition cannot be inferred from constraints or observable behaviour alone. Instead, cognition emerges through the mediation of identity tension structures that organise interpretation and decision formation.

\begin{figure}[h]
    \centering
    \includegraphics[width=0.95\linewidth]{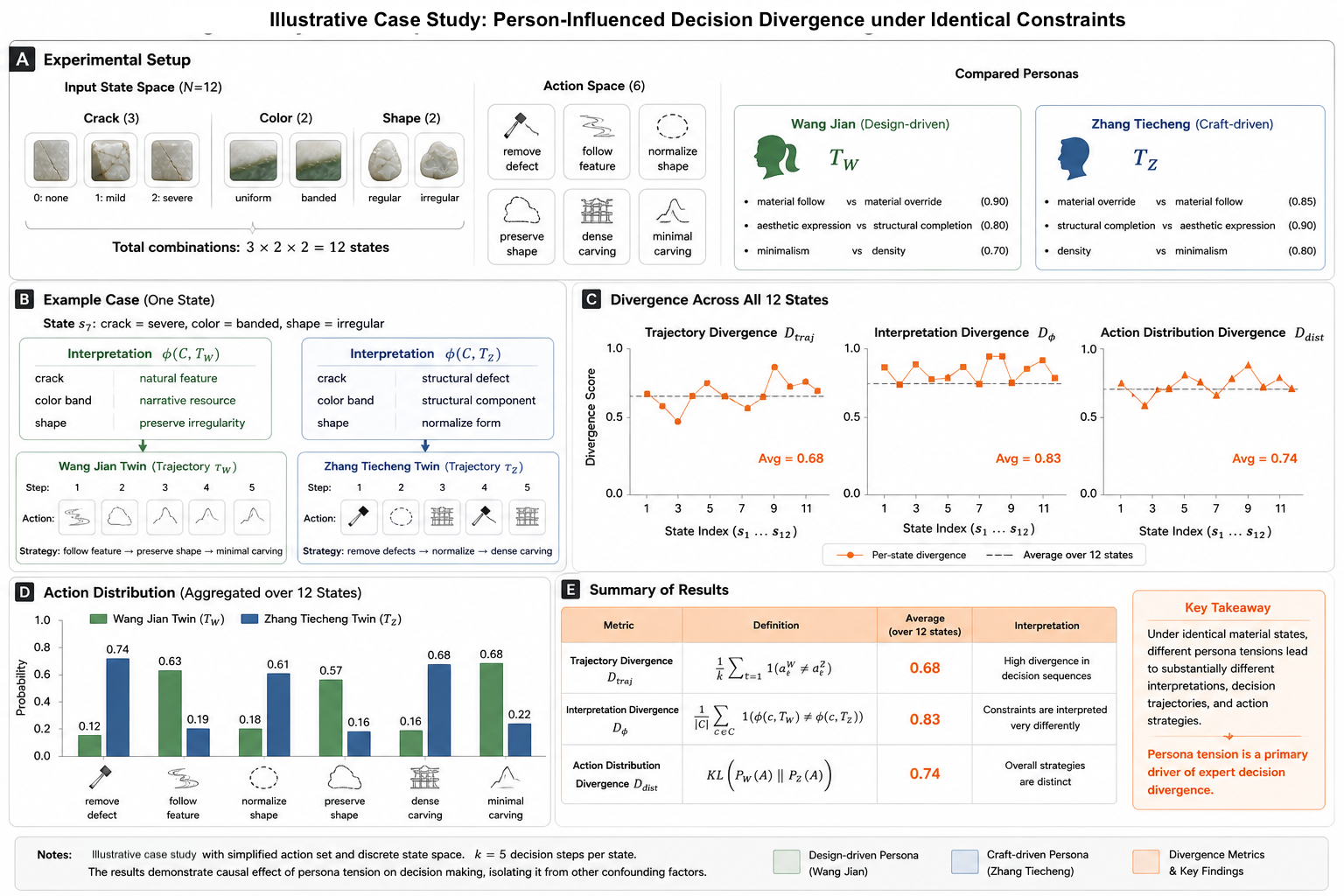}
    \caption{Illustrative case study demonstrating persona-induced decision divergence under identical constraints. Under identical material states, different identity tension structures produce divergent interpretations, decision trajectories, and action distributions, illustrating the central mechanism proposed by EICM.}
    \label{fig:illustrative-case-study}
\end{figure}

To operationalise this claim, we construct an illustrative case study designed to isolate the cognitive effects of persona-conditioned identity tensions under identical material conditions. The purpose of the case study is not empirical benchmarking or predictive evaluation. Rather, it serves as a computational demonstration of the core EICM hypothesis: identical constraints may produce divergent interpretations, decision trajectories, and action distributions when mediated by different identity tension structures.

Traditional jade carving is used as the illustrative domain because it requires continuous interpretation of uncertain material conditions and therefore provides a suitable environment for observing identity-mediated cognition.

Figure~\ref{fig:illustrative-case-study} demonstrates persona-induced decision divergence under identical constraints. Under identical material states, different identity tension structures produce divergent interpretations, decision trajectories, and action distributions, illustrating the central mechanism proposed by EICM.

Figure~\ref{fig:illustrative-case-study} operationalises the central theoretical claim of EICM through an illustrative case study designed to isolate the cognitive effects of identity tension structures under identical constraints.

Unlike empirical benchmarking experiments, the purpose of the case study is conceptual rather than predictive. The case study demonstrates how identical material states may generate divergent interpretations, decision trajectories, and action distributions when mediated by different persona-conditioned tension structures.

The case study controls for external constraints while varying internal identity orientations represented by contrasting expert personas. As a result, divergence emerges not from differences in available information or environmental conditions, but from differences in interpretive mediation.

Within the framework, interpretation is represented as a transformation function:

\begin{equation}
\phi(C,T)
\end{equation}

where situational constraints $C$ acquire cognitive meaning through identity tension structures $T$.

The resulting divergence across trajectories and action distributions illustrates a core implication of EICM: expert cognition cannot be reconstructed through behavioural observation alone because observable actions are downstream manifestations of latent identity-mediated cognitive organisation.

Accordingly, Figure~\ref{fig:illustrative-case-study} serves not as empirical validation, but as a computational demonstration of how EICM may be operationalised as an interpretable framework for modelling expert cognition beyond behaviour.

\subsection{Case Study Setup}

Figure~\ref{fig:illustrative-case-study} presents the overall case study structure. The case study defines a simplified material state space composed of three constraint dimensions:

\begin{equation}
S = \{s_{crack}, s_{color}, s_{shape}\}
\end{equation}

where the crack condition belongs to $\{\text{none}, \text{mild}, \text{severe}\}$, the colour structure belongs to $\{\text{uniform}, \text{banded}\}$, and the shape structure belongs to $\{\text{regular}, \text{irregular}\}$.

The resulting state space contains:

\begin{equation}
3 \times 2 \times 2 = 12
\end{equation}

distinct material states.

A simplified action space is also defined:

\begin{equation}
A = \{\text{remove defect}, \text{follow feature}, \text{normalise shape}, \text{preserve shape}, \text{dense carving}, \text{minimal carving}\}
\end{equation}

These actions represent stylised carving strategies rather than literal procedural operations.

Two contrasting expert personas are then introduced.

\paragraph{Persona $T_W$: Design-Driven Orientation.}
This persona prioritises material following, expressive interpretation, openness, and minimalism. The corresponding identity tensions emphasise dialogue with material irregularity rather than structural normalisation.

\paragraph{Persona $T_Z$: Craft-Driven Orientation.}
This persona prioritises structural completion, material control, density, and technical precision. The corresponding identity tensions emphasise stability, completeness, and compositional normalisation.

Importantly, both personas operate under identical material conditions. The only manipulated variable is the internal tension structure represented by persona orientation.

\subsection{Interpretation Divergence}

Within EICM, constraints do not possess fixed cognitive meaning independent of interpretation. Instead, meaning emerges through identity-mediated tension structures.

The case study operationalises interpretation as:

\begin{equation}
\phi(C,T)
\end{equation}

where $C$ represents constraints, $T$ represents persona-conditioned identity tension structures, and $\phi$ represents interpretive transformation.

Figure~\ref{fig:illustrative-case-study} illustrates one representative state:

\begin{equation}
s_7 = (\text{crack} = \text{severe}, \text{color} = \text{banded}, \text{shape} = \text{irregular})
\end{equation}

Under the design-driven persona $T_W$, cracks are interpreted as natural features, colour bands become narrative resources, and irregularity is preserved as expressive asymmetry.

Under the craft-driven persona $T_Z$, cracks are interpreted as structural defects, colour transitions become compositional components, and irregularity becomes a problem requiring normalisation.

The same constraint state therefore, produces fundamentally different interpretive representations depending on the identity tension structure organising cognition. This divergence demonstrates the central EICM assumption that constraints alone cannot determine expert reasoning.

\subsection{Trajectory Divergence}

Interpretive divergence subsequently produces different decision trajectories. Within the framework, action trajectories are represented as:

\begin{equation}
\tau = \pi(S, \phi(C,T))
\end{equation}

where $\phi(C,T)$ defines interpretation and $\pi$ defines the resulting decision policy.

Under the design-driven pathway, the decision trajectory prioritises following material features, preserving irregularity, and minimal intervention. Under the craft-driven pathway, the trajectory prioritises defect removal, structural normalisation, and dense compositional completion.

Although both trajectories originate from identical constraints, they diverge because interpretation itself has already diverged through persona-conditioned tension structures.

To quantify this phenomenon, the case study measures trajectory divergence across all 12 states:

\begin{equation}
D_{traj} = \frac{1}{k}\sum_{t=1}^{k}\mathbf{1}(a_t^W \neq a_t^Z)
\end{equation}

The case study produces an average trajectory divergence score of:

\begin{equation}
D_{traj} = 0.68
\end{equation}

indicating substantial divergence in decision sequences despite identical material inputs.

\subsection{Interpretation and Action Distribution Divergence}

Beyond individual trajectories, the case study further evaluates divergence at the interpretive and behavioural distribution levels.

Interpretation divergence is defined as:

\begin{equation}
D_{\phi} = \frac{1}{|C|}\sum_{c \in C}\mathbf{1}(\phi(c,T_W) \neq \phi(c,T_Z))
\end{equation}

The resulting score is:

\begin{equation}
D_{\phi} = 0.83
\end{equation}

This indicates that the two personas interpret constraints differently in most material states.

Action distribution divergence is additionally evaluated through aggregated policy differences:

\begin{equation}
D_{dist} = KL(P_W(A) \parallel P_Z(A))
\end{equation}

yielding:

\begin{equation}
D_{dist} = 0.74
\end{equation}

This result demonstrates that divergent identity tensions generate not only isolated decision differences, but systematically distinct behavioural distributions across contexts.

Importantly, the case study suggests that behaviour alone cannot reconstruct the underlying cognitive mechanism. Similar external conditions may produce stable yet fundamentally different action distributions depending on how constraints are interpreted through identity tensions.

\subsection{Identity Tension as a Latent Cognitive Variable}

The illustrative case study supports the central theoretical claim of EICM: identity tension functions as a latent cognitive variable mediating between constraints and action.

Existing behaviour-centric approaches often assume that cognition can be approximated through mappings between state and action. However, the present case study demonstrates that identical state spaces may produce divergent decision systems when mediated by different tension structures.

This implies that expert cognition cannot be adequately modeled through behavioural imitation alone. Instead, EICM proposes that cognition requires reconstruction of the latent interpretive structures through which experts organise meaning, prioritise constraints, and stabilise values.

Within the framework:

\begin{equation}
C \rightarrow T \rightarrow V \rightarrow A
\end{equation}

constitutes the minimal reconstructable pathway of expert cognition.

Accordingly, tension is not treated as incidental conflict or emotional variability, but as the primary organisational mechanism underlying expert reasoning divergence.

\subsection{Implications for Computational Expert modelling}

The case study provides several implications for future AI systems attempting to model expertise.

First, expert cognition appears to require representation of interpretive mediation rather than behavioural output alone. Observable actions are insufficient for reconstructing the internal structures generating expert reasoning.

Second, tacit knowledge may be partially operationalised through identity tension modelling. Rather than representing expertise as procedural rules alone, systems may require explicit modelling of competing commitments and value stabilisation dynamics.

Third, the framework suggests a transition from behaviour imitation towards identity-structured cognition reconstruction.

This shift has potential implications for expert digital twins, tacit knowledge preservation, human-centred AI systems, cultural cognition modelling, professional reasoning simulation, and adaptive expert tutoring systems.

Accordingly, EICM proposes that future expert intelligence systems may require computational representations of identity-mediated tension structures as foundational cognitive variables underlying decision formation.

\section{Methodological Implications}

The preceding sections position EICM as both a theoretical framework and a methodological orientation for modelling expert cognition. If expert reasoning emerges through the mediation of constraints, identity tensions, and stabilised values, then methods for modelling expertise must move beyond observable behaviour alone. This section outlines the methodological implications of EICM for computational expert modelling, tacit knowledge reconstruction, expert digital twins, and human-centred AI systems.

\subsection{From Behavioural Imitation to Cognitive Reconstruction}

A central implication of EICM is that expert modelling should move beyond behavioural imitation towards reconstruction of internal cognitive organisation.

Most existing AI systems model expertise primarily through observable outputs, including demonstrations, trajectories, preferences, or reward-aligned actions. Within such approaches, cognition is typically approximated through mappings between states and behaviour.

However, the preceding theoretical and illustrative case study analyses suggest that observable behaviour alone is insufficient for reconstructing expert reasoning. Identical constraints may produce divergent judgements when interpreted through different identity tension structures, while similar behaviours may emerge from fundamentally different cognitive organisations.

EICM therefore proposes a methodological shift:

\begin{equation}
\text{Behaviour} \rightarrow \text{Cognition}
\end{equation}

rather than:

\begin{equation}
\text{Behaviour} \approx \text{Cognition}
\end{equation}

Within this perspective, expert modelling becomes a process of reconstructing the latent interpretive structures that organise judgement, prioritisation, and value stabilisation.

\subsection{Identity Tension as a Computational Variable}

EICM introduces identity tension as a reconstructable computational variable for expert cognition modelling.

Conventional approaches frequently operationalise cognition through optimisation objectives, utility maximisation, preference estimation, or policy learning. In contrast, EICM proposes that cognition may instead be organised around competing identity commitments that shape interpretation under constraints.

This introduces a new modelling unit:

\begin{equation}
T = \{i_1, i_2, \ldots, i_n\}
\end{equation}

where identity commitments interact to produce structured tensions guiding interpretation and action formation.

Methodologically, this implies that future expert systems may require representations of:

\begin{itemize}
    \item competing professional commitments;
    \item aesthetic orientations;
    \item ethical priorities;
    \item symbolic reasoning structures; and
    \item long-term stylistic consistency.
\end{itemize}

Rather than treating these as secondary metadata, EICM positions them as primary cognitive variables underlying expert decision divergence.

\subsection{modelling Interpretation Rather Than Action Alone}

Existing behaviour-centric systems frequently treat interpretation as implicit or irrelevant. Inputs are mapped directly onto outputs without explicitly modelling how experts cognitively organise meaning.

EICM instead proposes interpretation as a primary reconstructable layer:

\begin{equation}
\phi(C,T)
\end{equation}

where constraints acquire cognitive significance through identity-mediated transformation.

This distinction is methodologically important because expert disagreement often emerges not from differences in information availability, but from differences in interpretive organisation.

Accordingly, future systems may require explicit modelling of:

\begin{itemize}
    \item interpretive framing;
    \item semantic prioritisation;
    \item symbolic association;
    \item contextual meaning assignment; and
    \item constraint re-representation.
\end{itemize}

Under this framework, cognition is not merely a response to conditions, but a structured process of meaning construction.

\subsection{towards Computational modelling of Tacit Knowledge}

Tacit knowledge has historically been difficult to formalise because much expert reasoning cannot be fully articulated as explicit procedural rules.

EICM suggests that tacit knowledge may be partially reconstructable through identity-tension structures rather than procedural decomposition alone. Within the framework, tacit cognition emerges through repeated stabilisation of interpretive negotiation processes under constraints. Long-term practice produces relatively stable value structures that subsequently organise expert consistency across contexts.

This perspective provides a possible pathway for computational modelling of:

\begin{itemize}
    \item professional intuition;
    \item aesthetic judgement;
    \item craft reasoning;
    \item situated expertise; and
    \item culturally embedded decision systems.
\end{itemize}

Rather than extracting isolated rules, future systems may instead reconstruct:

\begin{equation}
\text{Constraint} \rightarrow \text{Tension} \rightarrow \text{Value}
\end{equation}

as the minimal cognitive pathway underlying tacit expertise.

\subsection{Implications for Expert Digital Twins}

EICM has direct implications for the development of expert digital twins and cognition-centred AI systems.

Most current digital twin approaches reproduce external behavioural characteristics or procedural workflows. However, such systems often struggle to preserve long-term judgement consistency, stylistic coherence, tacit interpretation, or identity-dependent reasoning.

Within EICM, expert digital twins would require reconstruction of:

\begin{itemize}
    \item identity commitments;
    \item tension negotiation pathways;
    \item interpretive structures; and
    \item stabilised value organisations.
\end{itemize}

This implies a transition from behaviour-cloning systems towards identity-structured cognitive architectures capable of simulating expert reasoning under novel constraints.

Accordingly, future expert twins may function not merely as behavioural replicas, but as interpretable cognitive systems capable of reproducing characteristic modes of professional judgement.

\subsection{Human-centred AI and Interpretability}

EICM additionally contributes to ongoing discussions surrounding interpretability and human-centred AI.

Behaviour-centric systems frequently produce outputs without transparent representations of how constraints were interpreted internally. As a result, expert reasoning remains difficult to explain, audit, or align with human judgement structures.

By explicitly modelling:

\begin{equation}
C \rightarrow T \rightarrow V \rightarrow A
\end{equation}

EICM introduces a layered interpretive structure capable of exposing intermediate cognitive organisation.

This architecture may improve:

\begin{itemize}
    \item interpretability of expert systems;
    \item explanation of divergent decisions;
    \item alignment with professional reasoning; and
    \item transparency of value formation processes.
\end{itemize}

Importantly, EICM suggests that explainability should not be limited to action justification alone but should include reconstruction of the identity tensions organising interpretation itself.

\subsection{towards Identity-Structured AI Systems}

More broadly, EICM proposes a transition from behaviour-centric artificial intelligence towards identity-structured cognition systems.

Within this perspective, intelligence is not defined solely by task performance or optimisation efficiency, but by the capacity to organise interpretation through stable identity-mediated value structures.

Such systems may be capable of preserving tacit cultural reasoning, maintaining professional consistency, adapting under ambiguous constraints, and generating coherent long-term judgement trajectories.

Accordingly, EICM reframes expert AI not as systems that merely imitate behaviour, but as systems that reconstruct the latent cognitive organisations underlying expert reasoning processes.

This shift establishes a possible methodological foundation for future research in expert cognition modelling, tacit knowledge AI, cultural intelligence systems, expert digital twins, identity-centred reasoning architectures, and human--AI cognitive collaboration.

\section{Discussion}

This section discusses the broader theoretical implications of EICM for expert cognition modelling, tacit knowledge research, and future AI systems. Building on the framework and illustrative case study presented above, we argue that EICM offers a shift from modelling expertise as observable performance towards modelling expertise as identity-mediated cognitive organisation.

\subsection{Reframing Expert Cognition Beyond Behaviour}

This paper has argued that expert cognition cannot be adequately modeled through observable behaviour alone. Existing computational approaches frequently assume that expertise can be reconstructed through demonstrations, trajectories, or optimised decision outputs. While such approaches have achieved substantial success in highly structured environments, they remain limited in domains where cognition depends on tacit interpretation, identity negotiation, and long-term value consistency.

EICM proposes an alternative perspective in which expert cognition is modeled as an identity-structured interpretive process operating under situational constraints. Within this framework, constraints define the external problem space, tensions organise interpretive cognition, values stabilise judgement consistency, and behaviour emerges as the downstream manifestation of internal cognitive organisation.

This formulation shifts the primary object of expert modelling from action itself towardss the latent structures underlying expert reasoning.

\subsection{Theoretical Contribution of Identity Tension}

The central theoretical contribution of EICM lies in positioning identity tension as the core cognitive mechanism of expertise.

Conventional computational frameworks frequently interpret conflict through optimisation difficulty, uncertainty, or incomplete information. EICM instead proposes that many forms of expert reasoning emerge through negotiation between competing identity commitments.

This distinction is important because expert disagreement often persists even under comparable informational conditions. Such divergence cannot always be explained through knowledge deficits or environmental variation alone. EICM explains these phenomena by arguing that experts interpret constraints through different identity organisations, producing divergent reasoning trajectories and value stabilisations.

Accordingly, the framework reframes cognition not as direct environmental adaptation, but as structured identity-mediated interpretation.

\subsection{Implications for Tacit Knowledge Research}

The framework additionally contributes to longstanding discussions surrounding tacit knowledge and situated expertise.

Research on tacit knowledge has repeatedly emphasised that expert reasoning frequently exceeds explicit procedural articulation. However, many existing accounts remain descriptive and difficult to operationalise computationally.

EICM proposes that tacit expertise may become partially reconstructable through modelling identity tensions and value stabilisation processes. Within this perspective, tacit cognition is not treated as mysterious or inaccessible knowledge hidden beneath behaviour. Instead, it is conceptualised as a relatively stable interpretive organisation emerging through repeated negotiation of constraints under identity commitments.

This formulation offers a possible bridge between qualitative studies of expertise and computational cognition modelling.

\subsection{Limitations of the Current Framework}

Several limitations of the present work should be acknowledged.

First, EICM currently remains a theoretical and illustrative framework rather than a fully validated empirical model. The illustrative case study presented in Section 5 demonstrates operational plausibility rather than predictive performance.

Second, identity commitments and tension structures are difficult to formalise extensively. Real-world expert cognition may involve highly dynamic, multi-layered, and context-dependent identity interactions exceeding the simplified representations proposed here.

Third, the framework currently emphasises interpretive cognition more strongly than social or collaborative cognition. Many forms of expertise emerge through collective institutional, cultural, or organisational interaction that extends beyond individual identity structures.

Finally, the present formulation does not yet specify standardised procedures for extracting identity tensions from empirical data sources such as interviews, behavioural traces, or multimodal expert interaction records.

These limitations indicate that EICM should presently be understood as a foundational conceptual architecture rather than a completed computational system.

\subsection{Future Directions}

Several directions emerge for future research.

\paragraph{Empirical validation.}
Future work may investigate whether identity tension structures can be empirically reconstructed from expert interviews, longitudinal practice records, or multimodal decision traces.

\paragraph{Computational operationalisation.}
Further studies may develop formal computational representations of identity commitments, tension negotiation dynamics, interpretive transformation functions, and value stabilisation processes.

\paragraph{Expert digital twins.}
EICM may additionally provide a foundation for cognition-centred expert digital twins capable of reproducing long-term professional reasoning rather than isolated behavioural outputs alone.

\paragraph{Human--AI collaborative systems.}
Future systems may incorporate identity-aware reasoning architectures enabling AI systems to interact with experts through interpretable cognitive structures rather than purely statistical behaviour prediction.

\paragraph{Cross-domain expertise modelling.}
The framework may also be extended beyond craftsmanship domains towards medicine, design, education, law, scientific reasoning, and other forms of high-level professional cognition involving tacit judgement and identity-dependent interpretation.

\subsection{towards Identity-Structured Artificial Intelligence}

More broadly, EICM suggests a possible transition in AI research from behaviour-centric intelligence towards identity-structured cognition systems.

Current AI systems frequently prioritise behavioural competence, optimisation efficiency, and output prediction. While effective for many tasks, such systems may remain limited in their capacity to reproduce coherent expert judgement under ambiguity, conflicting commitments, or culturally embedded reasoning contexts.

EICM proposes that future expert intelligence systems may require explicit modelling of interpretive mediation, identity tension, value stabilisation, and long-term cognitive consistency.

Within this perspective, intelligence is not defined solely by successful task completion, but by the capacity to organise meaning, prioritise commitments, and maintain coherent reasoning structures across changing conditions.

Accordingly, EICM advances a foundational direction for modelling expert cognition beyond behaviour.

\section{Conclusion}

This paper introduced the \emph{Expert Identity Cognition Model} (EICM), a theoretical framework for modelling expert cognition beyond observable behaviour.

Existing approaches to expertise modelling have largely focused on behavioural outputs, optimisation performance, or preference approximation. While effective in highly structured environments, such approaches remain insufficient for explaining forms of expertise involving tacit judgement, identity-dependent interpretation, and long-term cognitive consistency.

EICM addresses this limitation by proposing that expert cognition should be understood as an identity-structured interpretive process operating under situational constraints. Within the framework, constraints define the external problem space, tensions organise interpretive cognition through competing identity commitments, values stabilise judgement structures, and behaviour emerges as the downstream manifestation of internal cognitive organisation.

Accordingly, the framework shifts the primary object of expert modelling from behavioural reproduction towardss reconstruction of latent identity-mediated cognitive structures.

A central contribution of the paper is the positioning of identity tension as the core cognitive mechanism underlying expert reasoning divergence. Rather than treating cognition as direct adaptation to external constraints, EICM argues that constraints acquire cognitive meaning only through identity-mediated interpretation. This formulation explains why experts operating under comparable conditions may produce fundamentally different judgements, trajectories, and action distributions despite possessing similar technical competence.

To operationalise the framework, the paper further introduced an illustrative case study demonstrating how divergent identity tension structures may generate distinct interpretations and behavioural trajectories under identical constraints. The case study illustrated that observable behaviour alone is insufficient for reconstructing the cognitive organisation underlying expert reasoning.

The framework additionally contributes to research on tacit knowledge, expert digital twins, human-centred AI, and computational cognition by proposing a reconstructable pathway linking:

\begin{equation}
\text{Constraint} \rightarrow \text{Tension} \rightarrow \text{Value} \rightarrow \text{Action}
\end{equation}

Rather than conceptualising expertise as procedural competence alone, EICM reframes expertise as the stabilisation of interpretive identity structures developed through long-term negotiation of constraints.

More broadly, the paper proposes a transition from behaviour-centric artificial intelligence towardss identity-structured cognition systems capable of modelling tacit interpretation, value formation, and coherent professional reasoning across changing contexts.

Accordingly, EICM establishes a foundational direction for future research on expert cognition modelling beyond behaviour.

\section*{Acknowledgements}
This research is supported by the University of Sydney Postgraduate Award.


\bibliographystyle{unsrtnat}
\bibliography{references}  

@article{hussein2017imitation,
  title={Imitation learning: A survey of learning methods},
  author={Hussein, Ahmed and Gaber, Mohamed Medhat and Elyan, Eyad and Jayne, Chrisina},
  journal={ACM Computing Surveys (CSUR)},
  volume={50},
  number={2},
  pages={1--35},
  year={2017},
  publisher={ACM New York, NY, USA}
}

@book{sutton1998reinforcement,
  title={Reinforcement learning: An introduction},
  author={Sutton, Richard S and Barto, Andrew G and others},
  volume={1},
  year={1998},
  publisher={MIT press Cambridge}
}

@article{christiano2017deep,
  title={Deep reinforcement learning from human preferences},
  author={Christiano, Paul F and Leike, Jan and Brown, Tom and Martic, Miljan and Legg, Shane and Amodei, Dario},
  journal={Advances in neural information processing systems},
  volume={30},
  year={2017}
}

@article{ouyang2022training,
  title={Training language models to follow instructions with human feedback},
  author={Ouyang, Long and Wu, Jeffrey and Jiang, Xu and Almeida, Diogo and Wainwright, Carroll and Mishkin, Pamela and Zhang, Chong and Agarwal, Sandhini and Slama, Katarina and Ray, Alex and others},
  journal={Advances in neural information processing systems},
  volume={35},
  pages={27730--27744},
  year={2022}
}

@article{simon1972theories,
  title={Theories of bounded rationality},
  author={Simon, Herbert A and others},
  journal={Decision and organization},
  volume={1},
  number={1},
  pages={161--176},
  year={1972},
  publisher={Amsterdam}
}

@book{suchman1987plans,
  title={Plans and situated actions: The problem of human-machine communication},
  author={Suchman, Lucille Alice},
  year={1987},
  publisher={Cambridge university press}
}

@article{hollan2000distributed,
  title={Distributed cognition: toward a new foundation for human-computer interaction research},
  author={Hollan, James and Hutchins, Edwin and Kirsh, David},
  journal={ACM Transactions on Computer-Human Interaction (TOCHI)},
  volume={7},
  number={2},
  pages={174--196},
  year={2000},
  publisher={ACM New York, NY, USA}
}

@article{polanyi1966logic,
  title={The logic of tacit inference},
  author={Polanyi, Michael},
  journal={Philosophy},
  volume={41},
  number={155},
  pages={1--18},
  year={1966},
  publisher={Cambridge University Press}
}

@book{dreyfus1986mind,
  title={Mind over machine},
  author={Dreyfus, Hubert and Dreyfus, Stuart E and Athanasiou, Tom},
  year={1986},
  publisher={Simon and Schuster}
}

@book{schon2017reflective,
  title={The reflective practitioner: How professionals think in action},
  author={Sch{\"o}n, Donald A},
  year={2017},
  publisher={Routledge}
}

@incollection{klein2011expert,
  title={Expert intuition and naturalistic decision making},
  author={Klein, Gary},
  booktitle={Handbook of intuition research},
  year={2011},
  publisher={Edward Elgar Publishing}
}

@article{nonaka2007knowledge,
  title={The knowledge-creating company},
  author={Nonaka, Ikujir{\=o} and Takeuchi, Hirotaka},
  journal={Harvard business review},
  volume={85},
  number={7/8},
  pages={162},
  year={2007}
}

@book{schatzki2001practice,
  title={The practice turn in contemporary theory},
  author={Schatzki, Theodore R and Knorr-Cetina, Karin and Von Savigny, Eike and others},
  volume={44},
  year={2001},
  publisher={Routledge London}
}

@book{lave1991situated,
  title={Situated learning: Legitimate peripheral participation},
  author={Lave, Jean and Wenger, Etienne},
  year={1991},
  publisher={Cambridge university press}
}

@book{mccall1966identities,
  title={Identities and interactions.},
  author={McCall, George J and Simmons, Jerry Laird},
  year={1966},
  publisher={Free Press}
}

@incollection{stryker2001traditional,
  title={Traditional symbolic interactionism, role theory, and structural symbolic interactionism: The road to identity theory},
  author={Stryker, Sheldon},
  booktitle={Handbook of sociological theory},
  pages={211--231},
  year={2001},
  publisher={Springer}
}

@article{markus1987dynamic,
  title={The dynamic self-concept: A social psychological perspective.},
  author={Markus, Hazel and Wurf, Elissa},
  journal={Annual review of psychology},
  year={1987},
  publisher={Annual Reviews}
}

@article{oyserman2009identity,
  title={Identity-based motivation: Implications for action-readiness, procedural-readiness, and consumer behavior},
  author={Oyserman, Daphna},
  journal={Journal of Consumer Psychology},
  volume={19},
  number={3},
  pages={250--260},
  year={2009},
  publisher={Elsevier}
}

@incollection{schwartz1992universals,
  title={Universals in the content and structure of values: Theoretical advances and empirical tests in 20 countries},
  author={Schwartz, Shalom H},
  booktitle={Advances in experimental social psychology},
  volume={25},
  pages={1--65},
  year={1992},
  publisher={Elsevier}
}

@incollection{friedman2013value,
  title={Value sensitive design and information systems},
  author={Friedman, Batya and Kahn Jr, Peter H and Borning, Alan and Huldtgren, Alina},
  booktitle={Early engagement and new technologies: Opening up the laboratory},
  pages={55--95},
  year={2013},
  publisher={Springer}
}

@article{friedman2017survey,
  title={A survey of value sensitive design methods},
  author={Friedman, Batya and Hendry, David G and Borning, Alan},
  journal={Foundations and Trends{\textregistered} in Human--Computer Interaction},
  volume={11},
  number={2},
  pages={63--125},
  year={2017},
  publisher={Emerald Publishing Limited}
}

@inproceedings{sanchez2026mental,
  title={Mental Models in Human-AI Interaction: Systematic Review of Empirical Methodologies and Guidelines},
  author={Sanchez, T{\'e}o and Vereschak, Oleksandra and Deroy, Ophelia},
  booktitle={Proceedings of the 31st International Conference on Intelligent User Interfaces},
  pages={663--682},
  year={2026}
}

@inproceedings{cha2023unlocking,
  title={Unlocking the tacit knowledge of data work in machine learning},
  author={Cha, Inha and Oh, Juhyun and Park, Cheul Young and Han, Jiyoon and Lee, Hwalsuk},
  booktitle={Extended abstracts of the 2023 CHI conference on human factors in computing systems},
  pages={1--7},
  year={2023}
}






\end{document}